\documentclass[lettersize,journal]{IEEEtran}
\usepackage{amsmath,amsfonts}
\usepackage{algorithmic}
\usepackage{algorithm}
\usepackage{array}
\usepackage[caption=false,font=normalsize,labelfont=sf,textfont=sf]{subfig}
\usepackage{textcomp}
\usepackage{stfloats}
\usepackage{url}
\usepackage{verbatim}
\usepackage{graphicx}
\usepackage{xcolor}
\usepackage{cite}
\usepackage{hyperref}
\usepackage[utf8]{inputenc}
\usepackage[utf8]{inputenc}

\usepackage[switch,pagewise,columnwise]{lineno}

\begin{document}

\title{CCAT: Magnetic Sensitivity Measurements of Kinetic Inductance Detectors}

\author{Benjamin J. Vaughan and Yuhan Wang, Cody J. Duell, Jason Austermann, James R. Burgoyne, Scott Chapman, Steve K. Choi, Abigail T. Crites, Eliza Gazda, Ben Keller, Michael D. Niemack, Darshan A. Patel, Anna Vaskuri, Eve M. Vavagiakis, Michael Vissers, Samantha Walker, Jordan Wheeler, Ruixuan(Matt) Xie

\thanks{© 2026 IEEE.  Personal use of this material is permitted.  Permission from IEEE must be obtained for all other uses, in any current or future media, including reprinting/republishing this material for advertising or promotional purposes, creating new collective works, for resale or redistribution to servers or lists, or reuse of any copyrighted component of this work in other works.}

\thanks{Benjamin J. Vaughan, Yuhan Wang, Cody J. Duell, Abigail T. Crites, Ben Keller Michael D. Niemack, Darshan A. Patel, Samantha Walker, and Eve M. Vavagiakis are all members of the Department of Physics at Cornell Univerisity

Jason Austermann, Anna Vaskuri, Michael Vissers, and Jordan Wheeler are all members of the National Institutes of Standards and Technology

Steve K. Choi and Eliza Gazda are members of the Department of Physics at University of California, Riverside

Scott Chapman is with the Department of Physics and Atmospheric
Science, Dalhousie University

James R. Burgoyne and Ruixuan(Matt) Xie are members of the Department of Physics and Astronomy of University of British Columbia

Eve M. Vavagiakis is a member of the Department of Physics at Duke University

}
}

\newcommand{\alchip}{Al $280$~GHz}
\newcommand{\tinchip}{TiN $280$~GHz}
\newcommand{\eorchip}{EoR-Spec}
\newcommand{\fyst}{\textit{FYST}}

\maketitle

\begin{abstract}
The CCAT Observatory is a ground-based submillimeter to millimeter experiment located on Cerro Chajnantor in the Atacama Desert, at an altitude of 5,600 meters. CCAT features the 6-meter Fred Young Submillimeter Telescope (FYST), which will cover frequency bands from 210 GHz to 850 GHz using its first-generation science instrument, Prime-Cam. The detectors used in Prime-Cam are feedhorn-coupled, lumped-element superconducting microwave kinetic inductance detectors (KIDs). The telescope will perform wide-area surveys at speeds on the order of degrees per second. During telescope operation, the KIDs are exposed to changes in the magnetic field caused by the telescope's movement through Earth's magnetic field and internal sources within the telescope. We present and compare measurements of the magnetic sensitivity of three different CCAT KID designs at 100 mK. The measurements are conducted in a dilution refrigerator (DR) with a set of room temperature Helmholtz coils positioned around the DR. We discuss the implications of these results for CCAT field operations.
\end{abstract}


\section{Introduction}
\label{sec:intro}

The Cerro Chajnantor Atacama Telescope (CCAT) \cite{CCAT_obs} is a ground-based submillimeter to millimeter experiment located at an altitude of 5,600 m in the Atacama Desert. CCAT features the 6 m Fred Young Submillimeter Telescope (FYST). Prime-Cam \cite{Vavagiakis_2018}, one of FYST's first-generation science instruments, will observe in the 210–850 GHz range using feedhorn-coupled, lumped-element superconducting microwave kinetic inductance detectors (KIDs). Prime-Cam can host up to seven instrument modules, in two general types. Broadband modules, named for their band center, use broadband, dual-polarization KID arrays. The spectroscopically-enabled modules are designed for imaging spectroscopy, such as the Epoch of Reionization Spectrometer (EoR-Spec) module \cite{Freundt_2024} fielding a Fabry-Perot Interferometer and 210 GHz to 420 GHz non-polarized KID arrays. Each module contains three detector arrays, designed for that module's scientific goal. The KID arrays are fabricated on 550 $\mu$m silicon-on-insulator wafers and operated at around 100 mK in individual instrument modules. We aim to deploy Prime-Cam with two broadband instrument modules centered at 280 \cite{duell2024ccatcomparisons280ghz, Vaskuri_2025, Choi_2022,vavagiakis_2022} and 350 GHz \cite{Huber_2024} in 2026, followed by an additional broadband module at 850 GHz \cite{Anthony_2025} and the EoR-Spec module.

During observations, the telescope will conduct wide-area surveys (5–20,000 deg²) at scan speeds of degrees per second. The survey area and scan strategy depends on the science targets and are being finalized at the time of writing. In these observations, the detectors are exposed to magnetic field variations from the telescope’s motion through Earth’s magnetic field and from internal sources within the telescope. Previous studies have shown that magnetic fields impact superconducting materials which will affect KID performance \cite{Liu_2022,kid_magnetometer,magnetometer,Flanigan_2016,vortices_in_two_types}. Here, we investigate the effect of an external DC magnetic field on CCAT KIDs.

\section{Experimental Set-up}
\label{sec:exp_app}

\begin{figure*}[h!]
    \centering
    \includegraphics[width=1\linewidth]{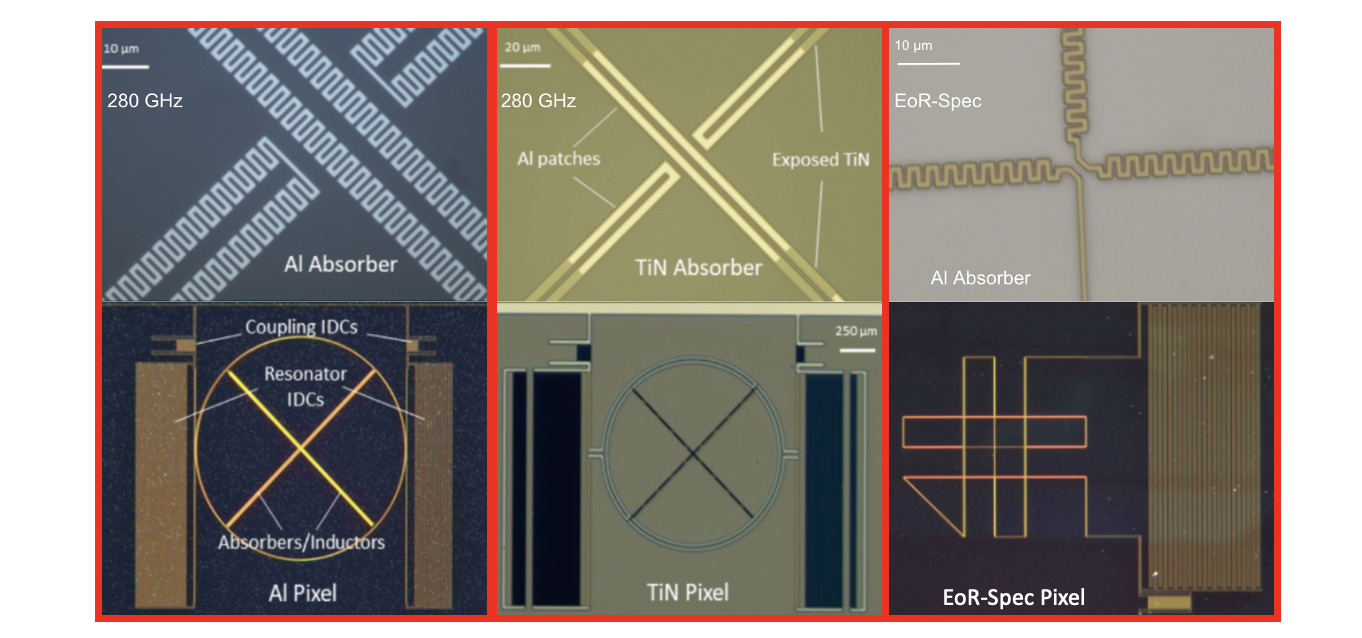}
    \caption{Al 280 GHz, TiN 280 GHz, and EoR-Spec pixel designs (bottom row) with close-ups of the respective absorbers (top row), shown from left to right. Figure adapted from Ref. \cite{duell2024ccatcomparisons280ghz}, with the EoR-Spec pixel design added.}
    \label{fig:det_designs}
\end{figure*}

\begin{figure}[h!]
    \centering
    \includegraphics[width=1\linewidth]{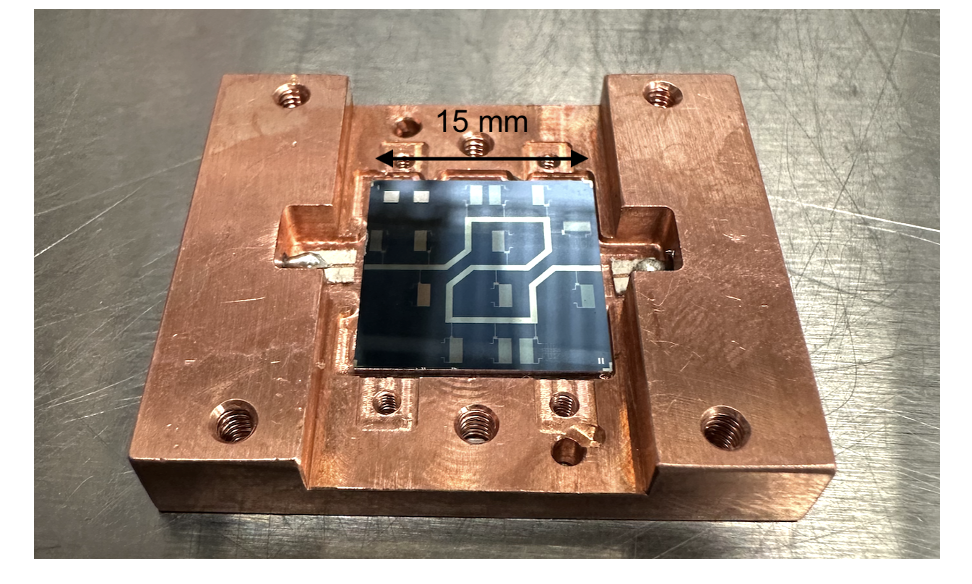}
    \caption{EoR-Spec test chip mounted in a copper test box and wire-bonded to microstrip pieces on the sides. The microstrip pieces contain copper traces and are soldered to SMP connectors on each side, not visible from this angle. The lid of the box is not shown. }
    \label{fig:detbox}
\end{figure}

For the magnetic sensitivity measurements, we used chips fabricated alongside the CCAT 280 GHz \cite{duell2024ccatcomparisons280ghz} and low-band 210–315 GHz EoR-Spec \cite{Freundt_2024} detector arrays with the same lithographic processes and materials. We refer to these as ‘witness chips.’ Three witness chips were packaged for this study: two 280 GHz chips (15 mm × 8 mm) made from aluminum (Al) and titanium-nitride (TiN), and one Al EoR-Spec chip (15 mm × 15 mm). The Al and TiN 280 GHz pixels (shown in the left and center panels of Figure \ref{fig:det_designs}) are lumped element polarimeter designs where one pixel has two absorbers measuring two orthogonal linear polarizations. Each detector is capacitively coupled to the transmission line by a coupling interdigitated capacitor (IDC). The resonance frequency ($f_{\rm r}$) is set by an inductor (with inductance L) which is also the absorber and parallel IDC (with capacitance C) as $f_{\rm r} = \frac{ 1 }{ 2 \pi \sqrt{L C} }$. The large difference in the kinetic inductance per square of the two materials led to a much finer linewidth and a meandered absorber for the Al detector, while the TiN absorber \cite{Vissers_2013} incorporated short patches of Al to tune detector responsivity\cite{duell2024ccatcomparisons280ghz}. While also having a lumped element design, the EoR-Spec detectors (shown in the right panel of Figure \ref{fig:det_designs}) do not require polarization sensitivity and use a meandered Hilbert curve absorber\cite{Li_2022}. Each chip was mounted in a separate, non-magnetic copper test box using rubber cement and read out via gold-plated brass SMA or SMP connectors with aluminum wire bonds. The box was sealed with a copper lid and brass screws. Figure \ref{fig:detbox} shows the EoR-Spec chip integrated into the base of the copper test box, as an example.

\begin{figure}[h!]
    \centering
    \includegraphics[width=1\linewidth]{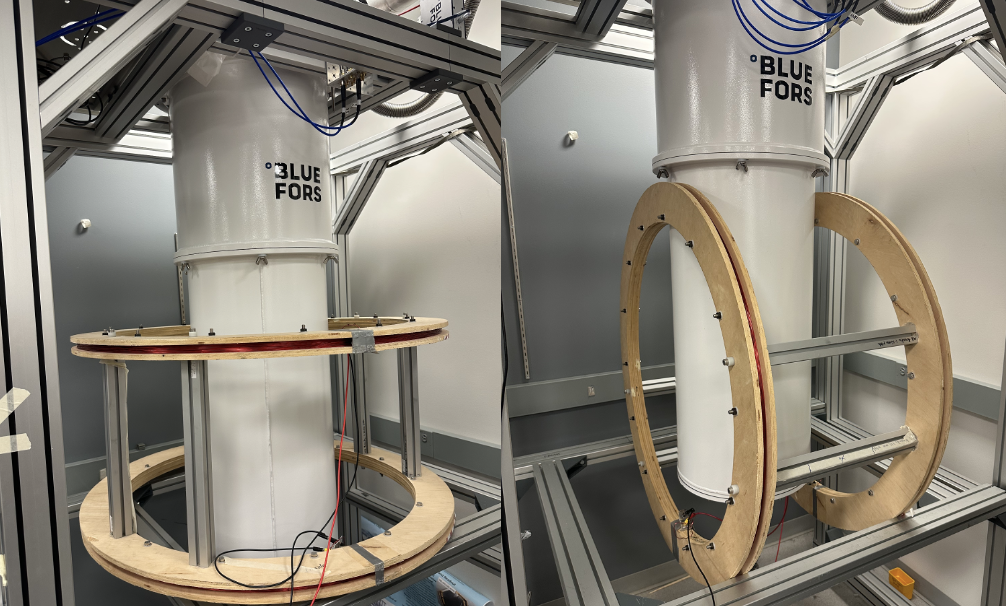}
    \caption{Helmholtz coils mounted outside the DR in two different configurations. \textit{Left}: field applied perpendicular to the detector absorber plane. \textit{Right}: parallel. The test chips are mounted on the DR mixing chamber near the coil center. The mu-metal magnetic shield used during cooldowns and the Lakeshore 460 gaussmeter are not shown. This mu-metal magnetic shield was held in place around the cryostat during cooldowns and removed once the cryostat was held at $100$~mK.}
    \label{fig:DR_coils}
\end{figure}

The cryogenic testing was performed in a Bluefors LD dilution refrigerator (DR). A set of Helmholtz coils was mounted on the DR frame outside the cryostat to provide an adjustable, uniform magnetic field as shown in Fig \ref{fig:DR_coils}. This set of coils, each with a radius of 44.6 cm and 267 turns of 16-AWG copper wire, with an approximately 45 cm separation between the two coils forming the Helmholtz pair, was previously used for similar measurements on other devices \cite{Vavagiakis_2021, Huber_2022}. A Lakeshore 460 three-channel gaussmeter was used to calibrate the coil performance by comparing calculated field strength at various drive voltages with measured values, and to confirm field uniformity at the center of the coil set. Additional measurements confirmed that the DR shells at various temperature stages do not provide additional magnetic shielding against the DC magnetic fields applied in this work.

All three witness chips were tested together, with their copper test boxes mounted on the DR mixing chamber (MXC) plate along the center of the Helmholtz coils. The TiN 280 GHz and EoR-Spec boxes were stacked on the top side of the mixing chamber plate. There is approximately 4 cm of vertical separation between the TiN and EoR-Spec chips, with the TiN chip on top. The Al 280 GHz box was mounted on the bottom of the mixing chamber plate, placing the Al chip approximately 5 cm below the EoR-Spec chip. This arrangement placed all boxes close to the center of the applied magnetic field and near each other to ensure they experienced the same field amplitude, in contrast to the dimensions of the Helmholtz coils. In two DR cooldowns, the Al 280 GHz box was flipped upside down to confirm that the measurements were independent of the chip’s upside-down orientation. Each chip is read out through an independent channel, similar to that planned for Prime-Cam \cite{Duell_thesis}, with cryogenic low-noise amplifiers at the 4 K stage on the RF output chain, various cryogenic attenuators on the RF input chain, and cryogenic coaxial cables. A custom-packaged Radio Frequency System-on-Chip (RFSoC) \cite{rfsoc, ccat_rfsoc} serves as the main room-temperature readout electronics, and an Agilent E5071C vector network analyzer was used to assist the measurements.

A room-temperature mu-metal magnetic shield was installed outside the DR during cooldowns, and removed once the cryostat was held at $100$~mK, before the magnetic measurements. Following each set of measurements, the magnetic shield was reinstalled while the detectors were warmed up above the critical temperature, $T_c$, of the superconducting aluminum (between 1.2 K and 1.4 K), before cooling back down. The importance of a shielded magnetic environment during cooldown past the KID’s critical temperature has been well documented in previous publications \cite{Flanigan_2016, Liu_2022} and is also discussed in Section \ref{sec:analysis}. After cooling to base temperature, the DR mixing chamber was servoed to 100 mK, using an active feedback (PID) loop that modulated heater power to maintain the temperature setpoint. The resonators from the test chips were individually tuned with different input RF powers to achieve high sensitivity while avoiding bifurcation \cite{Swenson_2013}. The following analysis uses all of the available detectors on each chip which amounts to five Al 280 GHz detectors, two TiN 280 GHz detectors, and five low-band 210–315 EoR-Spec detectors.

\begin{figure*}[ht!]
    \centering
    \includegraphics[width=1\linewidth]{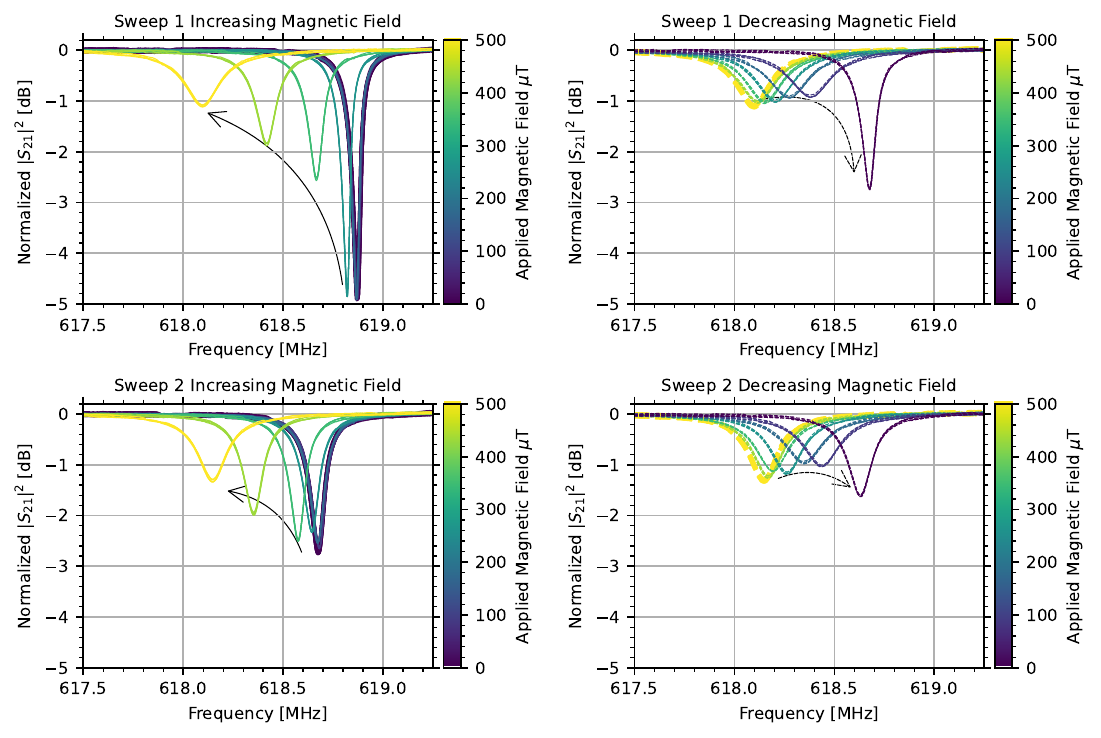}
     \caption{An example of how the \alchip\ resonator's frequency shifts in an external magnetic field. Each of the four panels represents a different sweep in magnetic field. In the top left we increase the magnetic from $0$-$500$~$\mu$T with the field pointing vertically upward (perpendicular to the plane of the pixels). After reaching $500$~$\mu$T we ramp back down to $0$~$\mu$T, as shown in the top right panel. The bottom two panels follow the same pattern, ramping up and down in magnetic field directly after the first sweep. The effect of the external magnetic on both the quality factor and the resonant frequency are clearly visible, as well as a degradation in the quality factor following each ramp that does not return to the nominal value, likely due to remnant magnetization.}
    \label{fig:S21_with_mag_field}
\end{figure*}

\section{Measurements and Analysis}
\label{sec:analysis}

To measure the magnetic sensitivity of the KIDs, the voltage across the coils was ramped up from $0$-$24$~V ($0$-$1$~A) in steps of $2$~V, this translated to a range of approximately $0$-$500$~$\mu$T in steps of ~$40$~$\mu$T. At each external magnetic field magnitude, $B$, we waited for ~$200$~seconds and then commenced a set of five scattering forward transmission ($S_{21}$) sweeps centered at each resonator's nominal resonance frequency, $f_0$. 

Figure \ref{fig:S21_with_mag_field} shows an example of the measured $S_{21}$ for a sample Al 280 GHz detector from two consecutive sweeps, where in each sweep the magnetic field was ramped up in magnitude and then back down. To better understand the  detectors' response to the external magnetic field, we extract the resonant frequency, $f_0$, and internal quality factor, $Q_i$,  by fitting the $S_{21}$ data to the following resonator model \cite{khalil_analysis_2012},
\begin{equation}\label{eq:full_s21}
    S_{21}(f) =\sqrt{ A_0f + A_1}e^{-j(2\pi f\tau + \phi)} \left(1 - \frac{Q_r}{Q_e}\frac{1}{1 + 2 j Q_r \frac{f - f_0}{f0}}\right).
\end{equation}
Above, the contributions from the readout network and cable are captured in $A_0$, $A_1$, the cable delay ($\tau$), and phase offset ($\phi$), while the resonator parameters $Q_r$, $Q_e$, and $f_0$, are the total quality factor, external coupling quality factor, and resonant frequency, respectively. The internal quality factor of the resonator is obtained from $Q_r$ and $Q_e$ by the following relation,
\begin{equation}
    \frac{1}{Q_r} = \text{Re}\left(\frac{1}{Q_e}\right) + \frac{1}{Q_i} = \frac{1}{Q_c}+ \frac{1}{Q_i}.
\end{equation}
$Q_c$ is the coupling quality factor, which is purely real and can be determined from $Q_e$ which is complex. Following an initial fit with the full set of frequency data to fix the cable components of the model, a much narrower window of data was then fit to a full nonlinear resonator model as described in \cite{Swenson_2013} and shown below. 
\begin{equation}
        S_{21} = 1 - \frac{Q_r}{Q_e}\frac{1}{1 + 2jy}.
\end{equation}
In this equation, $y$ is the resonator detuning in linewidths accounting for the non-linear kinetic inductance given by the solutions to the equation
\begin{equation}
    4y^3 - \left(Q_r\frac{f-f_0}{f_0}a+1\right)y = 0.
\end{equation}
Here, $a$ depicts the degree of non-linearity in the resonator. Multiple measurements were taken at each magnetic field strength. The values presented are the medians for the measurements. Error bars throughout the paper are defined as the measurement standard deviation and are plotted but too small to be visible. The temperature of the cryostat varied at the level of $\pm1$~mK throughout these sets of measurements. The gaussmeter precision was characterized through 20-minute magnetic field measurements at each voltage level. Due to uncertainties in the exact coil geometry and placement, as well as the precision of the gaussmeter, we estimate the precision of the applied magnetic field to be approximately $5$~$\mu$T. For clarity, the uncertainty on the magnetic field strength are not shown in any of these plots.

\begin{figure}
    \centering
    \includegraphics[width=1\linewidth]{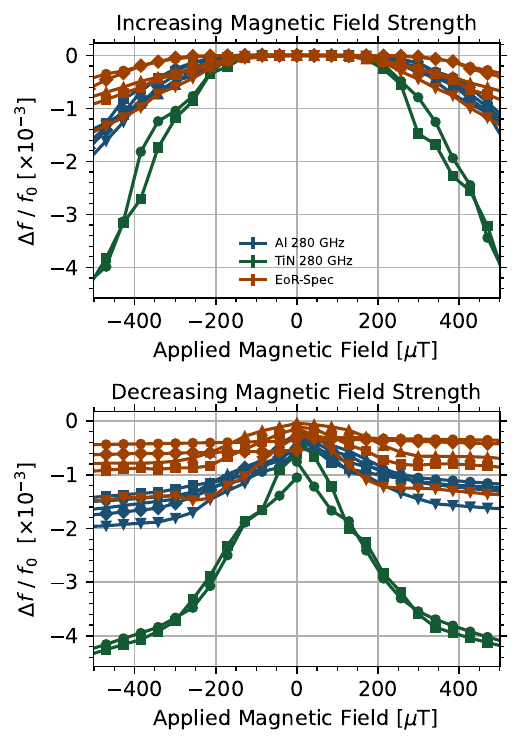}
    \caption{The fractional change in frequency shift for CCAT \alchip\ \tinchip\ and \eorchip detectors are plotted as a function of the applied external magnetic field. Different markers represent different detectors on a given test chips. In the top panel, the external magnetic field strength is increased from $0$ to $500$~$\mu$T in both field directions (via switching the direction of current through the coils), whereas, in the bottom panel the samples start at $500$~$\mu$T and the external field is brought back to $0$~$\mu$T. Shown here is that the resonators experience a hysteresis where the return curve has a different functional form than the initial ramp up in external magnetic field. }
    \label{fig:df_over_f}
\end{figure}

From these fits we obtain two main observables, the fractional change in resonant frequency, and the resonator's internal quality factor. We first investigate the fractional frequency shift via,
\begin{equation}
    \frac{\Delta f_0}{f_{0,z}} = \frac{f_0-f_{0,z}}{f_{0,z}}.
\end{equation}
Above, $f_{0,z}$ is the zero field resonant frequency. This comes from a set of target sweeps performed at $0$~$\mu$T, and while the cryostat was still enclosed by $\mu$-metal after reaching 100 mK for the first time.
Plotted in Figure \ref{fig:df_over_f} is $\Delta f / f_0$ versus $B$ over both vertical directions (positive with the field pointing vertically upward and negative downward). It is evident in Figure \ref{fig:df_over_f} that there is a hysteresis effect in which the resonator properties change differently when the magnetic field strength is increased compared to when it is decreased. This is further demonstrated in the quality factor of the resonator. In Figure \ref{fig:S21_with_mag_field}, the width and depth of the resonator changes as a function of magnetic field, and on subsequent sweeps the dip depth does not fully recover to its initial value of $5$~dB. 

\begin{figure}[h!]
    \centering
    \includegraphics[width=1\linewidth]{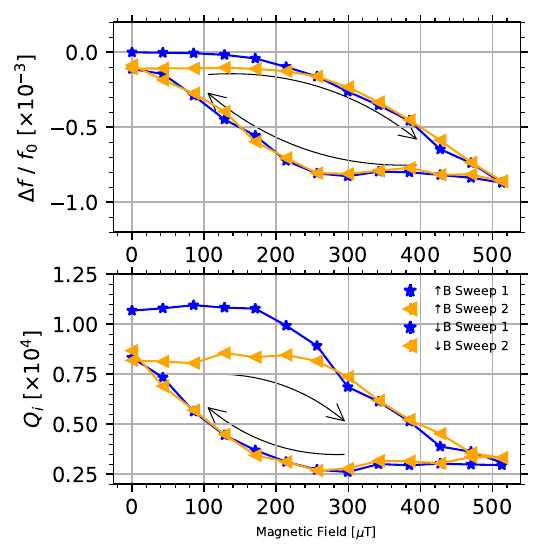}
    \caption{Using a characteristic resonator from the EoR-Spec witness chip we investigate the effect on the observables described in Section \ref{sec:analysis}. In each of the panels four curves are plotted representing two subsequent sweeps in magnetic field. The arrows in the plot indicate the direction of the change in $B$ during each sweep. A hysteresis effect is observed, where the response while decreasing the magnetic field to zero differs from that during the increasing field sweep, in both the fractional shifts of $f_0$ and the changes in $Q_i$. Furthermore, after the first magnetic field sweep, the observables do not return to their initial values. Subsequent sweeps return to values close to those observed in the second sweep but are not shown for clarity.}
    \label{fig:three_panel_figure}
\end{figure}

We investigate these effects in more detail by looking at an individual resonator that is characteristic of the whole set. Fractional shifts in $f_0$ and $Q_i$ are plotted for an EoR-Spec detector in Figure \ref{fig:three_panel_figure}. A similar hysteretic degradation of the quality factor is seen. We hypothesize that this is due to the formation of trapped magnetic flux or magnetic vortices within the superconducting films as they are exposed to these magnetic fields, introducing excess loss and decreasing the amount of stored energy available; however, further investigation is needed to confirm this.

\begin{figure}[h!]
    \centering
    \includegraphics[width=1\linewidth]{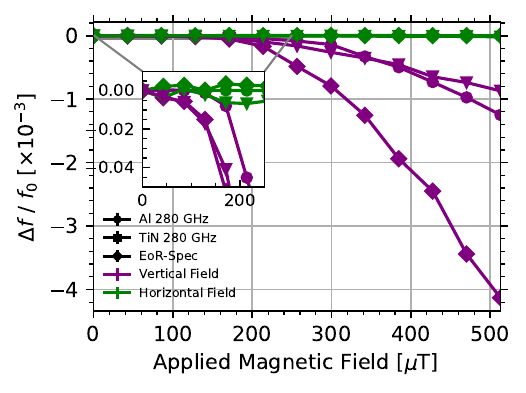}
    \caption{We investigate the effect of a magnetic field parallel to the absorber plane of the detectors via a comparison with the perpendicular field measurement. The solid green lines are the result of the parallel magnetic field measurement, whereas purple lines represent the perpendicular. One representative detector from each test chip is plotted here.  Across all three witness chips it is apparent that magnetic fields perpendicular to the plane of the detectors have the most impact. The inset provides a zoom-in on the low-field region of the applied magnetic field. The largest variation in Earth's field expected during telescope operation at the site is around 50 $\mu$T. We note that additional magnetic shielding has been implemented in the CCAT telescope receiver \cite{mod_cam_design}, and the impact of magnetic field variations encountered at the observation site is expected to be negligible, as further discussed in Section~\ref{sec:discussion}.}
    \label{fig:horizontal_field}
\end{figure}

We also study the effects of a magnetic field parallel to the detector absorber plane by placing the coils in a horizontal orientation. In the same manner as the other data-set we fit a non-linear resonator model to the data and extract the fractional shift in resonant frequency. As shown in Figure \ref{fig:horizontal_field}, the effects of the horizontal field on $\Delta f / f_0$ are a few orders of magnitude lower than that of the vertical orientation. 

\begin{figure}[h!]
    \centering
    \includegraphics[width=1\linewidth]{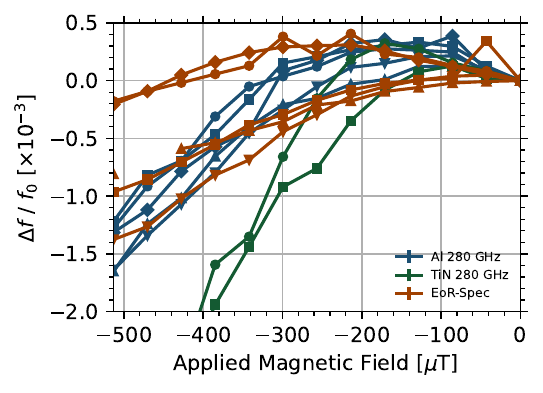}
    \caption{The fractional frequency shifts of the resonators as a function of the applied field when the magnetic shield was not in place while cooling down below the detectors' $T_c$ are plotted. A key difference between this Figure and the top panel of Figure \ref{fig:df_over_f} is the gradual increase and subsequent decrease in fractional frequency shift, visible as a bump at around $-200$~$\mu$T. }
    \label{fig:negative_bump}
\end{figure}

As an aside, we performed a set of measurements without the magnetic shield in place while cooling down below the detectors' $T_c$. We found that there is a bump in the $\Delta f / f_0$ when sweeping through the negative field, where the field direction points vertically downward as shown in Figure \ref{fig:negative_bump}. This observation emphasizes the importance of a shielded magnetic environment during cooldown as the temperature passes through $T_c$ and the chips become superconducting, as noted in previous publications \cite{Flanigan_2016, Liu_2022}

\section{Discussion}
\label{sec:discussion}

We find that, when subjected to external magnetic fields, the KIDs exhibit degraded internal quality factors and changes in responsivity, as quantified by $\Delta f / f_0$, with the effect dominated by magnetic fields perpendicular to the detector plane. We now discuss the implications this has on CCAT observations. 

First, we note that we were unable to decouple the hysteresis observed in the chip from hysteresis originating in the DR. Although the chip packaging was constructed from non-magnetic materials, components in the DR made of stainless steel, which are not feasible to replace, can still introduce magnetic hysteresis. Future studies of magnetic field effects on KIDs would benefit from either a cold Helmholtz coil that couples only to the test chip or a cryogenic magnetometer to monitor the local field around the test device.

At the observing site, the maximum expected magnitude of the Earth’s magnetic field is approximately 25 $\mu$T\footnote{Estimated using the NOAA magnetic field calculator at https://www.ngdc.noaa.gov/geomag/calculators/magcalc.shtml at the FYST location (22$^\circ$59$^\prime$09$^{\prime\prime}$S, 67$^\circ$44$^\prime$25$^{\prime\prime}$W).}. Changes in the magnetic field strength that the detectors experience due to the telescope moving through the Earth’s field should be less than twice this value. Taking the 50 $\mu$T as an upper bound to the possible change in field would yield a corresponding fractional frequency shift of less than $10^{-5}$, as shown in Figure~\ref{fig:horizontal_field}. Using the previously reported detector sensitivity for the Al 280 GHz detectors \cite{Vaskuri_2025}, under an optical loading of approximately 5.6 pW in the best weather quartile, this $10^{-5}$ fractional frequency shift corresponds to an optical signal of about 0.3 pW, which can be considered an extreme upper bound. 

Inside the telescope receiver, a Cryoperm A4K shield\footnote{Amuneal Mfg. Corp., Philadelphia, PA 19124} covers the back of the instrument modules and provides magnetic shielding at 4 K \cite{mod_cam_design}. Although the shielding factor has not been directly measured, simulations estimate it to be greater than 200, consistent with other experiments employing similar magnetic shielding schemes \cite{Ali_2020, Schillaci_2020}. Such a shielding factor would reduce a $25$~$\mu$T field to less than $0.125$~$\mu$T at the detector plane. Additionally, any shifts due to the magnitude field ought to occur as a long baseline common mode signal across all detectors in the module, meaning they will appear in the data in a manner similar to the atmospheric loading. Considering the relatively small changes in magnetic field magnitude, the module’s magnetic shielding, and the common-mode nature of the signal, we conclude that the magnetic field variations encountered while scanning through Earth’s field during telescope operations are expected to have a negligible impact on the detectors.

\section*{Acknowledgments}
The CCAT project, FYST and Prime-Cam instrument have been supported by generous contributions from the Fred M. Young, Jr. Charitable Trust, Cornell University, and the Canada Foundation for Innovation and the Provinces of Ontario, Alberta, and British Columbia. The construction of the FYST telescope was supported by the Gro{\ss}ger\"{a}te-program of the German Science Foundation (Deutsche Forschungsgemeinschaft, DFG) under grant INST 216/733-1 FUGG, as well as funding from Universität zu Köln, Universität Bonn and the Max Planck Institut für Astrophysik, Garching. The construction of EoR-Spec is supported by NSF grant AST-2009767. The construction of the 350 GHz instrument module for Prime-Cam is supported by NSF grant AST-2117631. S. Walker acknowledges support from the National Science Foundation under Award No. 2503181. S. Chapman acknowledges both the Natural Sciences and Engineering Research Council of Canada and the Canadian Foundation for Inovation.

\bibliographystyle{IEEEtran}

\bibliography{bare_jrnl_new_sample4}

@INPROCEEDINGS{CCAT_obs,
       author = {{Terry}, Herter and {Battaglia}, Nicholas and {Basu}, Kaustuv and {Beringue}, Benjamin and {Bertoldi}, Frank and {Chapman}, Scott and {Choi}, Steve and {Cothard}, Nicholas and {Chung}, Dongwoo and {Erler}, Jens and {Fich}, Michel and {Foreman}, Simon and {Gallardo}, Patricio and {Gao}, Jiansong and {Graf}, Urs and {Haynes}, Martha and {Herter}, Terry and {Hilton}, Gene and {Hubmayr}, Johannes and {Johnstone}, Doug and {Komatsu}, Eiichiro and {Magnelli}, Benjamin and {Mauskopf}, Phil and {McMahon}, Jeffrey and {Meerburg}, Daan and {Meyers}, Joel and {Mittal}, Avirukt and {Niemack}, Michael and {Nikola}, Thomas and {Parshley}, Stephen and {Riechers}, Dominik and {Stacey}, Gordon and {Stutzki}, Juergen and {Vavagiakis}, Eve and {Viero}, Marco and {Vissers}, Michael},
        title = "{The CCAT-Prime Submillimeter Observatory}",
    booktitle = {Bulletin of the American Astronomical Society},
         year = 2019,
       volume = {51},
        month = sep,
          eid = {213},
        pages = {213},
          doi = {10.48550/arXiv.1909.02587},
archivePrefix = {arXiv},
       eprint = {1909.02587},
 primaryClass = {astro-ph.IM},
       adsurl = {https://ui.adsabs.harvard.edu/abs/2019BAAS...51g.213T}
}

@inproceedings{ccat_rfsoc,
author = {James R. Burgoyne and Adrian K. Sinclair and Scott C. Chapman and Steve K. Choi and Cody J. Duell and Anthony I. Huber and Zachary B. Huber and Ben Keller and Lawrence Lin and Michael D. Niemack and Douglas Scott and Eve M. Vavagiakis and Samantha Walker and Matt Xie},
title = {{CCAT: FYST prime-cam readout software: a framework for massively scalable KID arrays}},
volume = {13101},
booktitle = {Software and Cyberinfrastructure for Astronomy VIII},
editor = {Jorge Ibsen and Gianluca Chiozzi},
organization = {International Society for Optics and Photonics},
publisher = {SPIE},
pages = {131013C},
year = {2024},
doi = {10.1117/12.3019028},
URL = {https://doi.org/10.1117/12.3019028}
}

@article{Swenson_2013,
    author = {Swenson, L. J. and Day, P. K. and Eom, B. H. and Leduc, H. G. and Llombart, N. and McKenney, C. M. and Noroozian, O. and Zmuidzinas, J.},
    title = {Operation of a titanium nitride superconducting microresonator detector in the nonlinear regime},
    journal = {Journal of Applied Physics},
    volume = {113},
    number = {10},
    pages = {104501},
    year = {2013},
    month = {03},
    issn = {0021-8979},
    doi = {10.1063/1.4794808},
    url = {https://doi.org/10.1063/1.4794808},
    eprint = {https://pubs.aip.org/aip/jap/article-pdf/doi/10.1063/1.4794808/15108284/104501\_1\_online.pdf},
}

@INPROCEEDINGS{rfsoc,
       author = {{Sinclair}, Adrian K. and {Stephenson}, Ryan C. and {Roberson}, Cody A. and {Weeks}, Eric L. and {Burgoyne}, James and {Huber}, Anthony I. and {Mauskopf}, Philip M. and {Chapman}, Scott C. and {Austermann}, Jason E. and {Choi}, Steve K. and {Duell}, Cody J. and {Fich}, Michel and {Groppi}, Christopher E. and {Huber}, Zachary and {Niemack}, Michael D. and {Nikola}, Thomas and {Rossi}, Kayla M. and {Sriram}, Adhitya and {Stacey}, Gordon J. and {Szakiel}, Erik and {Tsuchitori}, Joel and {Vavagiakis}, Eve M. and {Wheeler}, Jordan D.},
        title = "{CCAT-prime: RFSoC based readout for frequency multiplexed kinetic inductance detectors}",
    booktitle = {Millimeter, Submillimeter, and Far-Infrared Detectors and Instrumentation for Astronomy XI},
         year = 2022,
       editor = {{Zmuidzinas}, Jonas and {Gao}, Jian-Rong},
       series = {Society of Photo-Optical Instrumentation Engineers (SPIE) Conference Series},
       volume = {12190},
        month = aug,
          eid = {121900W},
        pages = {121900W},
          doi = {10.1117/12.2629722},
archivePrefix = {arXiv},
       eprint = {2208.07465},
 primaryClass = {astro-ph.IM},
       adsurl = {https://ui.adsabs.harvard.edu/abs/2022SPIE12190E..0WS}
}

@INPROCEEDINGS{mod_cam_design,
       author = {{Vavagiakis}, Eve M. and {Duell}, Cody J. and {Austermann}, Jason and {Beall}, James and {Bhandarkar}, Tanay and {Chapman}, Scott C. and {Choi}, Steve K. and {Coppi}, Gabriele and {Dicker}, Simon and {Devlin}, Mark and {Freundt}, Rodrigo G. and {Gao}, Jiansong and {Groppi}, Christopher and {Herter}, Terry L. and {Huber}, Zachary B. and {Hubmayr}, Johannes and {Johnstone}, Doug and {Keller}, Ben and {Kofman}, Anna M. and {Li}, Yaqiong and {Mauskopf}, Philip and {McMahon}, Jeff and {Moore}, Jenna and {Murphy}, Colin C. and {Niemack}, Michael D. and {Nikola}, Thomas and {Orlowski-Scherer}, John and {Rossi}, Kayla M. and {Sinclair}, Adrian K. and {Stacey}, Gordon J. and {Ullom}, Joel and {Vissers}, Michael and {Wheeler}, Jordan and {Xu}, Zhilei and {Zhu}, Ningfeng and {Zou}, Bugao},
        title = "{CCAT-prime: design of the Mod-Cam receiver and 280 GHz MKID instrument module}",
    booktitle = {Millimeter, Submillimeter, and Far-Infrared Detectors and Instrumentation for Astronomy XI},
         year = 2022,
       editor = {{Zmuidzinas}, Jonas and {Gao}, Jian-Rong},
       series = {Society of Photo-Optical Instrumentation Engineers (SPIE) Conference Series},
       volume = {12190},
        month = aug,
          eid = {1219004},
        pages = {1219004},
          doi = {10.1117/12.2630115},
archivePrefix = {arXiv},
       eprint = {2208.05468},
 primaryClass = {astro-ph.IM},
       adsurl = {https://ui.adsabs.harvard.edu/abs/2022SPIE12190E..04V}
}

@ARTICLE{vortices_in_two_types,
       author = {{Song}, C. and {Heitmann}, T.~W. and {Defeo}, M.~P. and {Yu}, K. and {McDermott}, R. and {Neeley}, M. and {Martinis}, John M. and {Plourde}, B.~L.~T.},
        title = "{Microwave response of vortices in superconducting thin films of Re and Al}",
      journal = {Phys. Rev. B},
         year = 2009,
        month = may,
       volume = {79},
       number = {17},
          eid = {174512},
        pages = {174512},
          doi = {10.1103/PhysRevB.79.174512},
archivePrefix = {arXiv},
       eprint = {0812.3645},
 primaryClass = {cond-mat.supr-con},
       adsurl = {https://ui.adsabs.harvard.edu/abs/2009PhRvB..79q4512S}
}

@ARTICLE{magnetometer,
       author = {{Luomahaara}, Juho and {Vesterinen}, Visa and {Gr{\"o}nberg}, Leif and {Hassel}, Juha},
        title = "{Kinetic inductance magnetometer}",
      journal = {Nature Communications},
         year = 2014,
        month = sep,
       volume = {5},
          eid = {4872},
        pages = {4872},
          doi = {10.1038/ncomms5872},
archivePrefix = {arXiv},
       eprint = {1401.0668},
 primaryClass = {physics.ins-det},
       adsurl = {https://ui.adsabs.harvard.edu/abs/2014NatCo...5.4872L}
}

@ARTICLE{kid_magnetometer,
       author = {{Levy-Bertrand}, F. and {Calvo}, M. and {Chowdhury}, U. and {Gomez}, A. and {Goupy}, J. and {Monfardini}, A.},
        title = "{Magnetic field tunable spectral response of kinetic inductance detectors}",
      journal = {Applied Physics Letters},
         year = 2025,
        month = jan,
       volume = {126},
       number = {4},
          eid = {042602},
        pages = {042602},
          doi = {10.1063/5.0231368},
archivePrefix = {arXiv},
       eprint = {2409.03356},
 primaryClass = {cond-mat.supr-con},
       adsurl = {https://ui.adsabs.harvard.edu/abs/2025ApPhL.126d2602L}
}

@article{Flanigan_2016,
    author = {Flanigan, D. and Johnson, B. R. and Abitbol, M. H. and Bryan, S. and Cantor, R. and Day, P. and Jones, G. and Mauskopf, P. and McCarrick, H. and Miller, A. and Zmuidzinas, J.},
    title = {Magnetic field dependence of the internal quality factor and noise performance of lumped-element kinetic inductance detectors},
    journal = {Applied Physics Letters},
    volume = {109},
    number = {14},
    pages = {143503},
    year = {2016},
    month = {10},
    issn = {0003-6951},
    doi = {10.1063/1.4964119},
    url = {https://doi.org/10.1063/1.4964119},
    eprint = {https://pubs.aip.org/aip/apl/article-pdf/doi/10.1063/1.4964119/14488594/143503\_1\_online.pdf},
}

@inproceedings{Liu_2022,
   title={Design and testing of kinetic inductance detector package for the Terahertz Intensity Mapper},
   url={http://dx.doi.org/10.1117/12.2629675},
   DOI={10.1117/12.2629675},
   booktitle={Millimeter, Submillimeter, and Far-Infrared Detectors and Instrumentation for Astronomy XI},
   publisher={SPIE},
   author={Liu, Lun-Jun and Janssen, Reinier M. J. and Bradford, Charles M. and Hailey-Dunsheath, Steven and Filippini, Jeffrey P. and Aguirre, James E. and Bracks, Justin S. and Corso, Anthony J. and Fu, Jianyang and Groppi, Christopher E. and Hoh, Jonathan R. and Keenan, Ryan P. and Lowe, Ian N. and Marrone, Daniel P. and Mauskopf, Philip D. and Nie, Rong and Redford, Joseph G. and Trumper, Isaac L. and Vieira, Joaquin D.},
   editor={Zmuidzinas, Jonas and Gao, Jian-Rong},
   year={2022},
   month=aug, pages={153} }

@inproceedings{Freundt_2024,
   title={CCAT: a status update on the EoR-Spec instrument module for Prime-Cam},
   url={http://dx.doi.org/10.1117/12.3018477},
   DOI={10.1117/12.3018477},
   booktitle={Millimeter, Submillimeter, and Far-Infrared Detectors and Instrumentation for Astronomy XII},
   publisher={SPIE},
   author={Freundt, Rodrigo G. and Li, Yaqiong and Henke, Doug W. and Austermann, Jason E. and Burgoyne, James R. and Chapman, Scott and Choi, Steve and Duell, Cody J. and Huber, Zachary B. and Niemack, Michael D. and Nikola, Thomas and Lin, Lawrence and Riechers, Dominik A. and Stacey, Gordon J. and Vaskuri, Anna K. and Vavagiakis, Eve M. and Wheeler, Jordan D. and Zou, Bugao},
   editor={Zmuidzinas, Jonas and Gao, Jian-Rong},
   year={2024},
   month=aug, pages={65} }

@inproceedings{Vavagiakis_2018,
   title={Prime-Cam: a first-light instrument for the CCAT-prime telescope},
   url={http://dx.doi.org/10.1117/12.2313868},
   DOI={10.1117/12.2313868},
   booktitle={Millimeter, Submillimeter, and Far-Infrared Detectors and Instrumentation for Astronomy IX},
   publisher={SPIE},
   author={Vavagiakis, Eve and Ahmed, Zeeshan and Ali, Aamir and Basu, Kaustuv and Battaglia, Nicholas and Bertoldi, Frank and Bond, Richard and Bustos, Ricardo and Chapman, Scott C. and Chung, Dongwoo and Cothard, Nicholas F. and Coppi, Gabriele and Dicker, Simon and Duell, Cody J. and Duff, Shannon M. and Erler, Jens and Fich, Michel and Gallardo, Patricio A. and Henderson, Shawn W. and Herter, Terry L. and Galitzki, Nicholas and Hilton, Gene and Hubmayr, Johannes and Irwin, Kent D. and Koopman, Brian J. and McMahon, Jeffrey and Murray, Norman and Niemack, Michael D. and Nikola, Thomas and Nolta, Michael and Orlowski-Scherer, John L. and Parshley, Stephen C. and Riechers, Dominik A. and Rossi, Kayla and Scott, Douglas and Sierra, Carlos and Silva-Feaver, Max and Simon, Sara M. and Stacey, Gordon J. and Stevens, Jason R. and Ullom, Joel N. and Vissers, Michael R. and Wollack, Edward J. and Xu, Zhilei and Zhu, Ningfeng and Walker, Samantha},
   editor={Zmuidzinas, Jonas and Gao, Jian-Rong},
   year={2018},
   month=jul, pages={64} }

@misc{duell2024ccatcomparisons280ghz,
      title={CCAT: Comparisons of 280 GHz TiN and Al Kinetic Inductance Detector Arrays}, 
      author={Cody J. Duell and Jason Austermann and James Beall and James R. Burgoyne and Scott C. Chapman and Steve K. Choi and Rodrigo G. Freundt and Jiansong Gao and Christopher Groppi and Anthony I. Huber and Zachary B. Huber and Johannes Hubmayr and Ben Keller and Yaqiong Li and Lawrence T. Lin and Justin Matthewson and Philip Mauskopf and Alicia Middleton and Colin C. Murphy and Michael D. Niemack and Thomas Nikola and Adrian K. Sinclair and Ema Smith and Jeff van Lanen and Anna Vaskuri and Eve M. Vavagiakis and Michael Vissers and Samantha Walker and Jordan Wheeler and Bugao Zou},
      year={2024},
      eprint={2406.06828},
      archivePrefix={arXiv},
      primaryClass={astro-ph.IM},
      url={https://arxiv.org/abs/2406.06828}, 
}

@article{Vaskuri_2025,
author = {Anna K. Vaskuri and Jordan D. Wheeler and Jason E. Austermann and Michael R. Vissers and James A. Beall and James Burgoyne and Victoria Butler and Scott Chapman and Steve K. Choi and Abigail Crites and Cody J. Duell and Rodrigo Freundt and Anthony Huber and Zachary B. Huber and Johannes Hubmayr and Jozsef Imrek and Ben Keller and Lawrence Lin and Alicia Middleton and Michael D. Niemack and Thomas Nikola and Douglas Scott and Adrian Sinclair and Ema Smith and Gordon Stacey and Joel Ullom and Jeffrey van Lanen and Eve M. Vavagiakis and Samantha Walker and Bugao Zou},
title = {{280-GHz aluminum MKID arrays for the Fred Young Submillimeter Telescope}},
volume = {11},
journal = {Journal of Astronomical Telescopes, Instruments, and Systems},
number = {2},
publisher = {SPIE},
pages = {026005},
year = {2025},
doi = {10.1117/1.JATIS.11.2.026005},
URL = {https://doi.org/10.1117/1.JATIS.11.2.026005}
}

@article{Choi_2022,
   title={CCAT-Prime: Characterization of the First 280 GHz MKID Array for Prime-Cam},
   volume={209},
   ISSN={1573-7357},
   url={http://dx.doi.org/10.1007/s10909-022-02787-9},
   DOI={10.1007/s10909-022-02787-9},
   number={5–6},
   journal={Journal of Low Temperature Physics},
   publisher={Springer Science and Business Media LLC},
   author={Choi, S. K. and Duell, C. J. and Austermann, J. and Cothard, N. F. and Gao, J. and Freundt, R. G. and Groppi, C. and Herter, T. and Hubmayr, J. and Huber, Z. B. and Keller, B. and Li, Y. and Mauskopf, P. and Niemack, M. D. and Nikola, T. and Rossi, K. and Sinclair, A. and Stacey, G. J. and Vavagiakis, E. M. and Vissers, M. and Tucker, C. and Weeks, E. and Wheeler, J.},
   year={2022},
   month=aug, pages={849–856} }

@ARTICLE{Huber_2022,
       author = {{Huber}, Zachary B. and {Li}, Yaqiong and {Vavagiakis}, Eve M. and {Choi}, Steve K. and {Connors}, Jake and {Cothard}, Nicholas F. and {Duell}, Cody J. and {Galitzki}, Nicholas and {Healy}, Erin and {Hubmayr}, Johannes and {Johnson}, Bradley R. and {Keller}, Ben and {McCarrick}, Heather and {Niemack}, Michael D. and {Wang}, Yuhan and {Xu}, Zhilei and {Zheng}, Kaiwen},
        title = "{The Simons Observatory: Magnetic Shielding Measurements for the Universal Multiplexing Module}",
      journal = {Journal of Low Temperature Physics},
         year = 2022,
        month = nov,
       volume = {209},
       number = {3-4},
        pages = {667-676},
          doi = {10.1007/s10909-022-02875-w},
archivePrefix = {arXiv},
       eprint = {2111.11495},
 primaryClass = {astro-ph.IM},
       adsurl = {https://ui.adsabs.harvard.edu/abs/2022JLTP..209..667H}
}

@ARTICLE{Vavagiakis_2021,
       author = {{Vavagiakis}, Eve M. and {Ahmed}, Zeeshan and {Ali}, Aamir and {Arnold}, Kam and {Austermann}, Jason and {Bruno}, Sarah Marie and {Choi}, Steve K. and {Connors}, Jake and {Cothard}, Nicholas and {Dicker}, Simon and {Dober}, Brad and {Duff}, Shannon and {Fanfani}, Valentina and {Healy}, Erin and {Henderson}, Shawn and {Ho}, Shuay-Pwu Patty and {Hoang}, Duc-Thuong and {Hilton}, Gene and {Hubmayr}, Johannes and {Krachmalnicoff}, Nicoletta and {Li}, Yaqiong and {Mates}, John and {McCarrick}, Heather and {Nati}, Federico and {Niemack}, Michael and {Silva-Feaver}, Maximiliano and {Staggs}, Suzanne and {Stevens}, Jason and {Vissers}, Mike and {Ullom}, Joel and {Wagoner}, Kasey and {Xu}, Zhilei and {Zhu}, Ningfeng},
        title = "{The Simons Observatory: Magnetic Sensitivity Measurements of Microwave SQUID Multiplexers}",
      journal = {IEEE Transactions on Applied Superconductivity},
         year = 2021,
        month = aug,
       volume = {31},
       number = {5},
          eid = {TASC.2021},
        pages = {TASC.2021},
          doi = {10.1109/TASC.2021.3069294},
archivePrefix = {arXiv},
       eprint = {2012.04532},
 primaryClass = {astro-ph.IM},
       adsurl = {https://ui.adsabs.harvard.edu/abs/2021ITAS...3169294V}
}

@phdthesis{Duell_thesis,
author="Duell, Cody",
year={2024},
title={Superconducting Microwave Resonators for Cosmology and Astrophysics With CCAT
},


language={English},

school = {Cornell University}
}

@inproceedings{Li_2022,
author = {Yaqiong Li and Jason Austermann and James Beall and Steve K. Choi and Cody J. Duell and Jiansong Gao and Zachary B. Huber and Johannes Hubmayr and Ben Keller and Lawrence T. Lin and Michael D. Niemack and Thomas Nikola and Gordon J. Stacey and Joel Ullom and Eve M. Vavagiakis and Michael Vissers and Jordan Wheeler and Bugao Zou},
title = {{CCAT-prime: the design of the epoch of reionization spectrometer detector arrays}},
volume = {12190},
booktitle = {Millimeter, Submillimeter, and Far-Infrared Detectors and Instrumentation for Astronomy XI},
editor = {Jonas Zmuidzinas and Jian-Rong Gao},
organization = {International Society for Optics and Photonics},
publisher = {SPIE},
pages = {121902G},
year = {2022},
doi = {10.1117/12.2630209},
URL = {https://doi.org/10.1117/12.2630209}
}

@ARTICLE{Anthony_2025,
  author={Huber, Anthony I. and Austermann, Jason and Beall, James A. and Burgoyne, James and Chapman, Scott and Henke, Douglas and Hubmayr, Johannes and Van Lanen, Jeffrey and Sinclair, Adrian and Vaskuri, Anna K. and Vissers, Michael R. and Wheeler, Jordan},
  journal={IEEE Transactions on Applied Superconductivity}, 
  title={High-Density Photon-Noise-Limited Multi-Octave Submillimeter Kinetic Inductance Detectors for the Prime-Cam 850 GHz Module}, 
  year={2025},
  volume={35},
  number={5},
  pages={1-7},
  doi={10.1109/TASC.2024.3518461}}

@article{Vissers_2013,
   title={Proximity-coupled Ti/TiN multilayers for use in kinetic inductance detectors},
   volume={102},
   ISSN={1077-3118},
   url={http://dx.doi.org/10.1063/1.4804286},
   DOI={10.1063/1.4804286},
   number={23},
   journal={Applied Physics Letters},
   publisher={AIP Publishing},
   author={Vissers, Michael R. and Gao, Jiansong and Sandberg, Martin and Duff, Shannon M. and Wisbey, David S. and Irwin, Kent D. and Pappas, David P.},
   year={2013},
   month=jun }

@INPROCEEDINGS{vavagiakis_2022,
       author = {{Vavagiakis}, Eve M. and {Duell}, Cody J. and {Austermann}, Jason and {Beall}, James and {Bhandarkar}, Tanay and {Chapman}, Scott C. and {Choi}, Steve K. and {Coppi}, Gabriele and {Dicker}, Simon and {Devlin}, Mark and {Freundt}, Rodrigo G. and {Gao}, Jiansong and {Groppi}, Christopher and {Herter}, Terry L. and {Huber}, Zachary B. and {Hubmayr}, Johannes and {Johnstone}, Doug and {Keller}, Ben and {Kofman}, Anna M. and {Li}, Yaqiong and {Mauskopf}, Philip and {McMahon}, Jeff and {Moore}, Jenna and {Murphy}, Colin C. and {Niemack}, Michael D. and {Nikola}, Thomas and {Orlowski-Scherer}, John and {Rossi}, Kayla M. and {Sinclair}, Adrian K. and {Stacey}, Gordon J. and {Ullom}, Joel and {Vissers}, Michael and {Wheeler}, Jordan and {Xu}, Zhilei and {Zhu}, Ningfeng and {Zou}, Bugao},
        title = "{CCAT-prime: design of the Mod-Cam receiver and 280 GHz MKID instrument module}",
    booktitle = {Millimeter, Submillimeter, and Far-Infrared Detectors and Instrumentation for Astronomy XI},
         year = 2022,
       editor = {{Zmuidzinas}, Jonas and {Gao}, Jian-Rong},
       series = {Society of Photo-Optical Instrumentation Engineers (SPIE) Conference Series},
       volume = {12190},
        month = aug,
          eid = {1219004},
        pages = {1219004},
          doi = {10.1117/12.2630115},
archivePrefix = {arXiv},
       eprint = {2208.05468},
 primaryClass = {astro-ph.IM},
       adsurl = {https://ui.adsabs.harvard.edu/abs/2022SPIE12190E..04V}
}

@INPROCEEDINGS{Huber_2024,
       author = {{Huber}, Zachary B. and {Lin}, Lawrence T. and {Vavagiakis}, Eve M. and {Freundt}, Rodrigo G. and {Butler}, Victoria and {Chapman}, Scott C. and {Choi}, Steve K. and {Crites}, Abigail T. and {Duell}, Cody J. and {Gallardo}, Patricio A. and {Huber}, Anthony I. and {Keller}, Ben and {Middleton}, Alicia and {Niemack}, Michael D. and {Nikola}, Thomas and {Orlowski-Scherer}, John and {Smith}, Ema and {Stacey}, Gordon and {Walker}, Samantha and {Zou}, Bugao},
        title = "{CCAT: Prime-Cam optics overview and status update}",
    booktitle = {Millimeter, Submillimeter, and Far-Infrared Detectors and Instrumentation for Astronomy XII},
         year = 2024,
       editor = {{Zmuidzinas}, Jonas and {Gao}, Jian-Rong},
       series = {Society of Photo-Optical Instrumentation Engineers (SPIE) Conference Series},
       volume = {13102},
        month = aug,
          eid = {1310222},
        pages = {1310222},
          doi = {10.1117/12.3020373},
archivePrefix = {arXiv},
       eprint = {2407.20873},
 primaryClass = {astro-ph.IM},
       adsurl = {https://ui.adsabs.harvard.edu/abs/2024SPIE13102E..22H}
}

@article{khalil_analysis_2012,
	title = {An analysis method for asymmetric resonator transmission applied to superconducting devices},
	volume = {111},
	issn = {0021-8979, 1089-7550},
	url = {http://arxiv.org/abs/1108.3117},
	doi = {10.1063/1.3692073},
	language = {en},
	number = {5},
	urldate = {2019-05-16},
	journal = {Journal of Applied Physics},
	author = {Khalil, M. S. and Stoutimore, M. J. A. and Wellstood, F. C. and Osborn, K. D.},
	month = mar,
	year = {2012},
	note = {arXiv: 1108.3117},
	pages = {054510},
}

@article{Schillaci_2020,
   title={Design and Performance of the First BICEP Array Receiver},
   volume={199},
   ISSN={1573-7357},
   url={http://dx.doi.org/10.1007/s10909-020-02394-6},
   DOI={10.1007/s10909-020-02394-6},
   number={3–4},
   journal={Journal of Low Temperature Physics},
   publisher={Springer Science and Business Media LLC},
   author={Schillaci, A. and Ade, P. A. R. and Ahmed, Z. and Amiri, M. and Barkats, D. and Thakur, R. Basu and Bischoff, C. A. and Bock, J. J. and Boenish, H. and Bullock, E. and Buza, V. and Cheshire, J. and Connors, J. and Cornelison, J. and Crumrine, M. and Cukierman, A. and Dierickx, M. and Duband, L. and Fatigoni, S. and Filippini, J. P. and Hall, G. and Halpern, M. and Harrison, S. and Henderson, S. and Hildebrandt, S. R. and Hilton, G. C. and Hui, H. and Irwin, K. D. and Kang, J. and Karkare, K. S. and Karpel, E. and Kefeli, S. and Kovac, J. M. and Kuo, C. L. and Lau, K. and Megerian, K. G. and Moncelsi, L. and Namikawa, T. and Nguyen, H. T. and O’Brient, R. and Palladino, S. and Precup, N. and Prouve, T. and Pryke, C. and Racine, B. and Reintsema, C. D. and Richter, S. and Schmitt, B. L. and Schwarz, R. and Sheehy, C. D. and Soliman, A. and Germaine, T. St. and Steinbach, B. and Sudiwala, R. V. and Thompson, K. L. and Tucker, C. and Turner, A. D. and Umiltà, C. and Vieregg, A. G. and Wandui, A. and Weber, A. C. and Wiebe, D. V. and Willmert, J. and Wu, W. L. K. and Yang, E. and Yoon, K. W. and Young, E. and Yu, C. and Zhang, C.},
   year={2020},
   month=feb, pages={976–984} }

@article{Ali_2020,
   title={Small Aperture Telescopes for the Simons Observatory},
   volume={200},
   ISSN={1573-7357},
   url={http://dx.doi.org/10.1007/s10909-020-02430-5},
   DOI={10.1007/s10909-020-02430-5},
   number={5–6},
   journal={Journal of Low Temperature Physics},
   publisher={Springer Science and Business Media LLC},
   author={Ali, Aamir M. and Adachi, Shunsuke and Arnold, Kam and Ashton, Peter and Bazarko, Andrew and Chinone, Yuji and Coppi, Gabriele and Corbett, Lance and Crowley, Kevin D. and Crowley, Kevin T. and Devlin, Mark and Dicker, Simon and Duff, Shannon and Ellis, Chris and Galitzki, Nicholas and Goeckner-Wald, Neil and Harrington, Kathleen and Healy, Erin and Hill, Charles A. and Ho, Shuay-Pwu Patty and Hubmayr, Johannes and Keating, Brian and Kiuchi, Kenji and Kusaka, Akito and Lee, Adrian T. and Ludlam, Michael and Mangu, Aashrita and Matsuda, Frederick and McCarrick, Heather and Nati, Federico and Niemack, Michael D. and Nishino, Haruki and Orlowski-Scherer, John and Sathyanarayana Rao, Mayuri and Raum, Christopher and Sakurai, Yuki and Salatino, Maria and Sasse, Trevor and Seibert, Joseph and Sierra, Carlos and Silva-Feaver, Maximiliano and Spisak, Jacob and Simon, Sara M. and Staggs, Suzanne and Tajima, Osamu and Teply, Grant and Tsan, Tran and Wollack, Edward and Westbrook, Bejamin and Xu, Zhilei and Zannoni, Mario and Zhu, Ningfeng},
   year={2020},
   month=apr, pages={461–471} }

\newpage

\vfill

\end{document}